\documentclass[12pt,preprint]{aastex} 
\slugcomment{} 

\newcommand\kms{\ifmmode {\rm km\ s}^{-1} \else km s$^{-1}$\fi}
\newcommand\eflux{\ifmmode {\rm ergs\ s}^{-1}\;{\rm cm}^{-2} \else  
	ergs s$^{-1}$ cm$^{-2}$\fi}  
\newcommand\phflux{\ifmmode {\rm photons\ s}^{-1}\;{\rm cm}^{-2} 
	\else  	photons s$^{-1}$ cm$^{-2}$\fi}  
\newcommand\ergsec{\ifmmode {\rm ergs\ s}^{-1} \else  
	ergs s$^{-1}$\fi}
\newcommand\Msun{\ifmmode M_{\odot} \else $M_{\odot}$\fi}

\shorttitle{NGC3516 - complex absorption} 
\shortauthors{Turner et al. 2004}

\begin{document}
\title{Complex X-ray  Absorption and the Fe K$\alpha$ Profile in NGC 3516} 

\author{ T.\ J.\ Turner\altaffilmark{1,2},  
 S.\ B.\ Kraemer\altaffilmark{3,4}, I.\ M.\ George\altaffilmark{1,2}, 
J.\ N.\ Reeves\altaffilmark{2,5}, M.\ C.\ Bottorff\altaffilmark{6}
}

\altaffiltext{1}{Joint Center for Astrophysics, Physics Department, 
University of Maryland, Baltimore County, 
1000 Hilltop Circle, Baltimore, MD 21250}
\altaffiltext{2} {Laboratory for High Energy Astrophysics, Code 662, 
	NASA/GSFC, Greenbelt, MD 20771}
\altaffiltext{3}{Department of Physics, 
The Catholic University of America, NASA/GSFC, Code 681, NASA/GSFC 
Greenbelt, MD 20771}
\altaffiltext{4}{Laboratory for Astronomy and Solar Physics, Code 681, 
NASA/GSFC, Greenbelt, MD 20771}  
\altaffiltext{5}{Universities Space Research Association, 7501 Forbes Blvd, 
Suite 206, Seabrook,  MD 20706-2253}
\altaffiltext{6}{Southwestern University, 1001 E. University Ave.,
Georgetown, TX 78626}

\begin{abstract}

We present data  from simultaneous {\it Chandra},    
{\it XMM-Newton} and {\it BeppoSAX} 
observations of the Seyfert 1 galaxy NGC 3516, 
taken during 2001 April and November. We have investigated the 
nature of the very flat observed X-ray spectrum. 
{\it Chandra} grating data show the presence of 
X-ray absorption lines, revealing two distinct components
of the absorbing gas, one which is consistent with our previous
model of the UV/X-ray absorber while the other, which is
outflowing at a velocity of 
$\sim 1100$ km s$^{-1}$, has a larger column density and is much more
highly ionized. The broad-band spectral characteristics of 
the X-ray continuum observed
with {\it XMM} during 2001 April, reveal the presence of a {\it third} 
layer of absorption consisting of a very large column
($\approx$ 2.5 x 10$^{23}$ cm$^{-2}$) of highly
ionized gas with a covering fraction $\sim$ 50\%. This low covering 
fraction suggests that 
the absorber lies within  a few lt-days of the 
X-ray source and/or is filamentary in
structure. Interestingly, these absorbers are not in thermal equilibrium with one another. 
The two new components are too highly ionized to be radiatively accelerated,
which we suggest is evidence for a hydromagnetic origin for the outflow. Applying our model to the
November dataset, we can account for the spectral variability 
primarily by a drop
in the ionization states of the absorbers, 
as expected by the change in the continuum flux. 
When this complex absorption is accounted for we find the 
underlying continuum to be typical of Seyfert 1 galaxies. 
The spectral curvature 
attributed to  the high column  absorber, in turn,  
reduces estimates of the flux and extent of any broad Fe emission 
line from the accretion disk.  
	\end{abstract}


	\section{Introduction}

It is now understood that intrinsic UV and X-ray absorption is present
in the spectra of at least half of all 
Seyfert 1 galaxies (\citealt{reyn97,geo98,cre1999}). High resolution 
UV spectra obtained with the spectrographs aboard the 
{\it Hubble Space Telescope} and, most recently, X-ray spectra
obtained with {\it Chandra} and {\it XMM} have revealed blueshifted
absorption lines, signifying massive outflows of gas from the active nuclei
of these galaxies (\citealt{cren03}, and references therein). 
The fact that intrinsic absorption is so common among Seyfert galaxies suggests
that the absorbers have large global covering factors and the inferred mass loss
rates are comparable to the mass accretion rates of the supermassive black holes (SMBH) that power the
active galactic nuclei (AGN). The X-ray and UV absorption is highly variable in a number of sources,
which may be the result of changes in ionization state (\citealt{bro85,cre00}) or transverse
motion (\citealt{cnk99,kra01a}). Although it has been suggested that the X-ray and UV absorption arises in the same 
gas \citep{mat94}, there is clearly a range of physical conditions in each component (e.g.,
\citealt{kk,kra01b,kaspiea02}). Several dynamical models have been proposed for the 
physical nature of mass outflows in AGN including Compton-heated winds \citep{beg83}, 
radiatively driven flows (e.g., \citealt{mur95,pro00}), hydromagnetic
(MHD) flows (e.g, \citealt{BP82,bott00}), and hybrids of the latter two (e.g., \citealt{knk94,deknb95,pro03});
most current models assume that the outflow originates at the accretion disk surrounding
the SMBH. 

NGC 3516 (z=0.008836; \citealt{keel96}) is one of the few Seyfert 1 galaxies
with UV absorption lines, specifically  N~{\sc v} $\lambda\lambda$ 1238.8, 1242.8, , C~{\sc iv}
$\lambda\lambda$ 1548.2, 1550.8, and Si~{\sc iv} $\lambda\lambda$ 1393.8, 1402.8, 
strong enough to have been detected with the
{\it International Ultraviolet Explorer} ({\it IUE}; \citealt{ulr83}). 
{\it IUE} monitoring campaigns found evidence for absorption line variations
on timescales as short as weeks (\citealt{vsb87,walt90,kol93}), and the equivalent
width of C~{\sc iv} appeared to be anti-correlated with the UV continuum flux, which
suggests that the absorber was responding to changes in the strength of the ionizing radiation.
This variability constrains the radial distance of the absorbers to be $\lesssim$ 10 pc 
(\citealt{vsb87,walt90}). Based on {\it Hubble
Space Telescope (HST)}/Goddard High Resolution Spectrograph 
observations, \citet{cre98} determined that the
C~{\sc iv} absorption lines consisted of four separate 
kinematic components: two relatively stable narrow components
near the core of the line, and two broad, variable blueshifted components. 

NGC 3516 also exhibits strong, highly variable X-ray absorption
(\citealt{kol93,np94,kriss96,mat97}). Based on their analysis of {\it ROSAT} observations, Mathur et al. (1997) suggested that the X-ray and UV absorption arose in the
same component, although \citet{kol93} argued that the X-ray absorption detected with {\it Ginga} consisted of at least 
two zones. Mathur et al. further argued that the absence of a highly blue-shifted broad UV absorption feature 
in {\it IUE} observations taken between
1989 and 1993 \citep{kor96} was due to the drop in density,
 hence increase in ionization state 
of the absorber as it moved radially outwards. Recent 
observations have found NGC~3516 
to be in a relatively low flux state. 
{\it Chandra} LETG data from 2000 October 6 suggested a model 
whereby a constant column of gas 
with $N_H \sim 8 \times 10^{21} {\rm atom\ cm^{-2}}$ 
reacts to changes in the nuclear ionizing flux \citep{netzer02}. Netzer et al. determined the radial distance of the 
absorber to be $\lesssim$ 6 x 10$^{17}$ cm (for H$_{0}$ $=$ 75 km s$^{-1}$ kpc$^{-1}$). Based on their analysis
of contemporaneous echelle spectra obtained with the 
{\it HST}/Space Telescope Imaging
Spectrograph, \citet{kraemer02} argued that the strong 
Ly$\alpha$, N~{\sc v}, and C~{\sc iv} lines arose
in the same column of gas. Furthermore, they concluded that the variability of the UV absorption was primarily
due to changes in the ionizing flux, whose effect is amplified for the more distant absorbers, which are 
ionized by radiation that has been filtered by intervening gas. This is clearly illustrated by the re-appearance
of the highly blue-shifted, broad UV absorption discussed by Mathur et al. With NGC 3516 in a low flux state, the combination of the weaker ionizing 
radiation and the higher opacity of the intervening gas induced such a large drop in ionization that
a component of very highly ionized gas recombined sufficiently to produce observable UV lines.

We present here {\it Chandra} HETG spectra, 
{\it XMM-Newton} (hereafter {\it XMM}) EPIC CCD spectra 
and  PDS data from simultaneous {\it BeppoSAX} observations  
of NGC 3516 from 2001 April and November. 
Previous analysis of a subset of these data concentrated on 
just the Fe K$\alpha$ regime in data from 2001 November. 
The Fe K$\alpha$ line showed interesting structure and rapid evolution 
during the 2001 November 
observation, and the combined HETG/EPIC data in the hard X-ray regime 
have been detailed  by \citet{turner02}. 
Initial analysis of the 2001 November data also revealed a very flat 
observed continuum shape.  November data from the  HETG  
showed no significant absorption lines, leaving ambiguity as to the 
nature of the flat observed spectrum that could be either 
intrinsically flat; have a 'normal' photon index and appear 
flat due to complex layers of absorption or be dominated by 
reprocessed components mimicing a flat spectrum 
(i.e. a very broad line from the accretion disk, hereafter denoted a 
'diskline',  of 
large equivalent width and associated Compton hump). As the ambiguity 
could not be broken using 2001 November data alone, \citet{turner02}
 used a simple 
flat powerlaw parameterization of those data, against which to examine the 
newly-discovered narrow and redshifted line emission from Fe. 

Analysis of data from 2001 April has now shown footprints of highly-ionized 
absorbing gas, whose presence had been suggested but never unambiguously confirmed before in this source.  
We present a detailed parameterization of  
NGC~3516 based on data from both April and November epochs of 2001.

\section{The Multi-satellite Observations}

\subsection{The {\it Chandra} Data}

{\it Chandra} observations of NGC~3516 were performed at two epochs 
during 2001 using the 
High-Energy Transmission Grating Spectrometer (HETGS; 
e.g. Marshall, Dewey \& Ishibashi 2004)
in conjunction with the `S-array' of the 
Advanced CCD Imaging Spectrometer (ACIS; 
e.g. Garmire et al 2003).
We note that to-date, these are the only {\it Chandra} 
observations of NGC~3516 to have been performed with the
HETGS/ACIS-S instrument configuration.
The first observation (obsids 2431 and 2080)
was performed over the period April 09 14:12 to April 11 14:53 UTC, 
but with an interruption (April 10 01:17--17:55 UTC)
when solar activity led to sufficiently high background levels 
that the detectors were turned off.
The second observation (obsid 2482) was performed between
2001 November 11 01:00 UTC to November 12 02:19 UTC.
From the times of the observation listed above, it 
can be seen that the second period of {\it Chandra}
observations in April ($\sim$21~hr during obsid 2080) 
is simultaneous with {\it XMM}, as is $\sim$10~hr of
the observation in November.

The {\it Chandra} data from both epochs were processed using the 
{\tt ciao} (v3.0.1) and {\tt ftools} (v5.3.1) software 
packages, along with calibration files from the 
{\tt CALDB} (v2.26) release\footnote{Including
the {\tt acisD1999-08-13contamN0003.fits} file to correct for 
the time-dependent contamination on the surface of the CCDs
(e.g. see Marshall et al. 2004).}.
We also made use of our own
software for a number of the analysis tasks.
The data were processed in the standard manner, 
including the removal of bad detector pixels and other detector
artifacts, removal of events with detector 
`grades' {\it not} equal to 0, 2, 3, 4, or 6.
The selection criteria used resulted in
exposures of $\sim$36 and 73~ks for the 
first and second segments of the April data, 
and 88~ks for the November observation.

The background was negligible and so no background subtraction was 
performed for the HETGS spectra.  
The mean count rates in the 0.7-8.0~keV band obtained for  
summed (positive and negative) 1$^{\rm st}$-orders of the 
Medium-Energy Grating (MEG) 
arm were 
$0.110\pm0.023$ and $0.211\pm0.027$ ct~s$^{-1}$
for the two segments 
of the April observations, and 
$0.123\pm0.013$ ct~s$^{-1}$ during the November observation.
The corresponding values for the High-Energy Grating (HEG) arm
were  $0.069\pm0.013$, 
$0.127\pm0.018$, and
$0.076\pm0.010$ ct~s$^{-1}$
(respectively).
In the spectral analysis presented below we 
consider  the data from obsid 2080, as this is simultaneous 
with the good subset of the PN observation. 

	\subsection{The {\it XMM} Data}

{\it XMM} observations  of NGC~3516 were performed covering 2001 
April 10  11:14 --  April 11  23:02  UTC
and November 09  23:13 -- November 11 10:54 UTC.    
EPIC data utilized the thin filter with PN Prime Small Window mode.  
These data were processed using SAS  v5.3.3. and  were screened to 
remove hot and bad pixels 
and periods of high background. For the April epoch we took 
only the subset of PN data which were simultaneous with 
the second piece of {\it Chandra}  coverage 
(April  10  17:56 -  14:53 UTC), yielding 
a 49 ks exposure. This is the period during which EPIC 
background levels were lowest, and data outside of this 
time range resulted in inadequate 
subtraction  of the background spectrum. 
In November data screening 
 yielded an exposure of 81 ks. 
At the flux level of NGC~3516  photon pileup is negligible 
in EPIC data.     
During the analysis we found the MOS spectra to have much lower 
signal-to-noise ratio than the PN data, 
thus we performed our analysis on the PN spectra alone. 
For the PN, instrument patterns  0 -- 4  were selected and   
spectra were extracted from a cell $\sim 0.94'$ diameter, 
centered on the source. Background spectra were extracted from a nearby 
region for the PN ($5 - 10$\% of the source count rate). 

NGC~3516 yielded 5.48$\pm0.01$ ct s$^{-1}$  during 2001 April and 
$\sim  3.25\pm0.01$ ct s$^{-1}$ during 2001 November, over the 
0.35 -- 10 keV band. Table~1 shows source fluxes for the two epochs, and 
for the earlier LETG observation which we pulled from the archives and 
compared to our best-fitting model  for extraction of fluxes.  
NGC~3516 occupied the lowest known X-ray flux state during the 2000 October 
and 2001 November epochs, and during the early part of the 2001 April 
{\it Chandra} observations (obsid2431), rising to a higher flux 
later in 2001 April. This time variability allowed us to 
track the source behavior with changing flux.  

\subsection{The {\it BeppoSAX} Data}

{\it BeppoSAX} observed NGC~3516 2001 
April 10 UT 16:05 -- April 12  22:48 UTC  
and 2001 November 09 UT 20:27 -- November 12 08:01 UTC.  
We utilized  data from 
 the Phoswich Detector System 
(PDS; \citealt{f97}) providing spectral data over 
$\sim 15$ -- 200 keV. 
We used products from the {\it BeppoSAX} database, created 
 using a standard data reduction using the SAXDAS software 
 version 2.0. 
Data from the four PDS 
phoswich units were combined, after gain equalization and linearization. 
The instruments were switched off during SAA passages and 
PDS data acquired during the first 5 minutes after each SAA 
passage were rejected to exclude periods of highly 
variable risetime threshold.  
Standard screening of the {\it BeppoSAX} data  yielded 
events files  with effective exposure times of 
40 ks (April) and 63 ks (November).

The  PDS had a collimated 
$fov$ with FWHM = 1.4 degrees. The region around NGC~3516 
is free of contaminating hard X-ray sources of significant flux 
within this radius. The PDS instrument was rocked 
back and forth to obtain source and 
background data covering the period of the observation. 
The PDS rocking mode 
generally provides a very reliable background subtraction. 
We obtained, after background subtraction,  
0.86$\pm0.04$ PDS cts $^{-1}$ during 2001 April;  
0.68$\pm0.03$ PDS cts $^{-1}$ during 2001 November.

\section {X-ray Spectral Analysis}

\subsection{ {\it Chandra} Evidence for Multiple Components of Absorption}

The HETGS data were 
compared to an  absorbed continuum model which parameterized the 
curvature of the source,  and 
then examined for residual features.  
In Fig.~1 we show some of the most interesting sections of 
the spectrum obtained during 2001 April, 
where we have combined the 
(flux-calibrated) $\pm1^{\rm st}$-order spectra from the 
MEG and HEG arms and plotted the data using 0.01\AA~bins
(corresponding to the $\sim 1\sigma$ width of the spectral 
resolution of the MEG, and twice that value for the HEG).
Immediately evident are several 
deep absorption lines from highly ionized species of several elements, 
most notably the H-like species of Mg, Si and S.

As discussed in Section 1, based on an analysis of a {\it Chandra} 
Low-Energy Transmission Grating Spectrometer (LETGS) spectrum of
 NGC 3516 obtained on 2000 October 6,  
\citet{netzer02} argued for the presence of a relatively low-ionization 
(log U $= -0.72$) absorber, with a column density log N$_{H}$ $=$ 21.9. 
\citet{kraemer02} suggested the same gas was the source of the strong UV 
absorption lines of H~{\sc i}, C~{\sc iv}, and N~{\sc v} detected in 
contemporaneous 
{\it HST/STIS} medium-resolution echelle spectra. Since the  
2001 April and November observations exhibited similar soft X-ray 
spectral characteristics,
it is plausible that a column of low ionization gas 
(which we will refer to as
the ``UV component'') was still present. 
Assuming a sufficiently large velocity dispersion within the
absorbing gas (we assume $300\ {\rm km\ s^{-1}}$ here),  
the ionization levels in such a component are sufficient to 
give rise to detectable absorption lines (due to inner-shell 
transitions) from Mg {\sc vii}, {\sc viii}, {\sc ix} and 
Si {\sc vii}, {\sc viii}, {\sc ix}.
However, this UV component is not sufficiently-ionized to 
contain any H- and He-like species of these elements.
Nevertheless, 
as can be seen from Fig.~1, we do see clear evidence for 
absorption lines from H-like Mg, Si and S. 
The presence of absorption lines from the He-like species 
is less clear, although the data are highly suggestive of 
a line due to Si~{\sc xiii}, and possibly also 
Mg~{\sc xi}.
Thus we have strong evidence of 
{\it at least} one other, much more highly-ionized, component of absorption.

In Fig.~2 we show the kinematical constraints that can be obtained 
from the Mg~{\sc xii}, Si~{\sc xiii}, Si~{\sc xiv}, and S~{\sc xvi}
lines (the He-like lines of Mg and S provided no useful 
constraints, primarily due to the signal-to-noise limitations).
For this analysis we adopted the rest-wavelengths quoted in 
\citet{ven96-lines},\citet{vea96}, which are in good agreement with
other compilations. Specifically for the H-like 
$1s\,^{2}S_{1/2} \rightarrow 2p\,^{2}P^{\rm o}_{3/2,1/2}$
doublets
(with oscillator strengths in the ratio 2:1) we assumed 
wavelengths of 
4.7274 \& 4.7328\AA\ for S~{\sc xvi}, 
6.1804 \& 6.1858\AA\ for Si~{\sc xiv}, and
8.4192 \& 8.4246\AA\ for Mg~{\sc xii}.
For the Si~{\sc xiii} 
$1s^2\,^{1}S_{0}  \rightarrow 1s.2p\, ^{1}P^{\rm o}_{1}$
transition we adopted 6.6480\AA.
For each ion, we constructed high-resolution 'opacity spectra'
for various combinations of ionic column density, $N$(ion),
and velocity dispersion, $v_{\sigma}$. 
 Fig.~3 shows 
the contours for ionic column density versus $v_{\sigma}$  
for these H-like lines, illustrating their significance at $>90\%$ 
confidence, and the constraints attainable on ionic column and 
velocity dispersion.

Both absorption-line (assuming full Voigt
profiles) and bound-free absorption were included in these 
grids, and we assumed that the absorbing material 
completely covers the underlying X-ray source (i.e. a
covering-factor of unity).
The grids were then used to construct model spectra by assuming 
an underlying continuum (see below).
These spectra were then convolved using the instrumental spectral
response and compared to the data using {\tt xspec} (v.11.3.1),
with the redshift of the 
opacity spectra 
(and hence the offset velocity, $v_{off}$) allowed to vary 
 with respect to that expected given the recessional velocity
of the host galaxy. 

For the analysis presented in Fig.~2, for each ion 
we restricted consideration to the data within 
$\pm10^4\ {\rm km\ s^{-1}}$ of the expected location 
(after correcting for the systemic velocity)
of the $1s \rightarrow 2p$ transition, and to within 
the same velocity range of the corresponding bound-free edge.
Given the limited signal-to-noise of the data in the 
vicinity of the $1s \rightarrow 2p$ lines, 
the data in the vicinity of the bound-free edge 
had to be included to prevent solutions in the 
low $v_{\sigma}$, very-high $N$(ion) region of 
parameter-space.
The MEG and HEG data within the above spectral ranges were fitted 
simultaneously (with the different instrumental responses 
taken into account). We adopted spectral bins of width 
0.010 and 0.005\AA\ for the MEG and HEG (respectively) and used a
C-statistic minimization technique within {\tt xspec}.
In all cases a 
simple, smooth, underlying continuum was assumed, the normalization 
of which was allowed to vary during the spectral fitting.

The dashed curves in Fig.~2 show the 90\% confidence regions in
$v_{\sigma}$--$v_{off}$ space obtained from the individual 
Mg~{\sc xii}, Si~{\sc xiii}, Si~{\sc xiv} and S~{\sc xvi} ions.
It can be seen that the results for the individual 
Mg~{\sc xii}, Si~{\sc xiii} and 
Si~{\sc xiv} ions are all consistent with 
$v_{off}$ in the range 
$-400 \lesssim v_{off} \lesssim -1400\ {\rm km\ s^{-1}}$, but 
$v_{\sigma}$ cannot be constrained.
The results for S~{\sc xvi} appear to be slightly discrepant
with $-1450 \lesssim v_{off} \lesssim -2050\ {\rm km\ s^{-1}}$,
and $v_{\sigma} \lesssim 400\ {\rm km\ s^{-1}}$.
This discrepancy may be partly the result of statistical fluctuations
in low signal-to-noise data
or may be indicative of additional spectral complexity 
caused by additional kinematic components (see below).
From a combined analysis of all four ions we 
found 
$v_{off} = -1140^{+310}_{-301}\ {\rm km\ s^{-1}}$. 
By comparison,  the UV absorption components show velocities 
up to $\sim 1300$ km/s \citep{kraemer02}.

In contrast,  
the mean outflow velocity from the lower ionization S, Si and  Mg lines was 
close to systemic, consistent with an origin within the UV absorbers. 
The strength of Si {\sc xi} line (see Fig.~1) may 
indicate the presence of an additional, distinct layer of gas. 
However, the relative weakness of Si{\sc x} and Si{\sc xii} features 
led us to question the significance of the Si{\sc xi} absorption line and 
thus we note it but did not attempt to model it. 

While {\it Chandra} grating spectra from 2001 April 
allowed detection of the discreet absorption 
lines, enabling us to pick out previously-unresolved layers of 
gas, there was still some ambiguity as to the physical 
conditions (i.e, column density and ionization) of each layer. 
The lines were not resolved by HETGS and thus we needed 
to fit the line profiles 
to obtain further  constraints on the gas.  There is a significant trade-off 
between the assumed turbulent velocity and the column density required to 
produce each line, as discussed in the next section. 

Unfortunately  the source was fainter during November and 
HETGS data from the 2001 November epoch showed no significant 
narrow absorption features 
 although those data are {\it consistent} 
with the presence of the absorption lines found in 2001 April.

\section{Modeling the Absorbers}

\subsection{Inputs to the Models}

Our general approach to modeling is to start with a single-component 
absorption-model, including 
additional layers only as necessary. Hence, although it is 
entirely likely that 
the intrinsic absorption is more complex, we initially 
restricted our models to include only the two components  
whose footprints are clearly seen in the HETG spectra. 
Our next step was to generate model tables of the two zones 
of X-ray absorbing gas, and then we used these to fit the moderate 
resolution EPIC PN spectra, using the good signal-to-noise 
available there to constrain the broad spectral components.  

The photoionization models for this study were generated using the Beta 5
version of Cloudy (G. Ferland 2003, private communication). We modeled the
absorbers as single-zoned slabs of atomic gas, irradiated by the 
central source. 
We assumed a similar spectral energy distribution of the continuum 
radiation emitted by the central engine to that described in 
\citet{kraemer02}, 
with the exception that we modified the spectral energy 
 index, $\alpha$, above 500 eV 
based on our initial
fitting of the 2001 April data, such that: $\alpha = -1.0$
for h$\nu$ $<$ 100 eV, $\alpha = -2.3$ over the range 100 eV $\leq$ h$\nu$
$<$ 500 eV, and $\alpha = -1.0$ above 500 eV. The models are parameterized in
terms of the dimensionless ionization parameter $U$, which is the ratio of
ionizing photons per H atom at the illuminated face of the slab. Based on our
models of the X-ray emission lines in NGC 3516 \citep{turner03}, we assumed
solar elemental abundances (e.g. \citealt{ga89})  with the 
exception of
nitrogen, for which we adopted an N/H ratio of 2.5 times the solar value. The
absorbing gas is assumed to be free of cosmic dust.

We initially generated a grid of photoionization models around
values of $U$ and N$_{H}$ suggested by the 
{\it Chandra} detections of the H- and He-like lines discussed above 
(which we will refer to as the ``High Ionization'' component). For 
simplicity, we assumed a 300 km s$^{-1}$
velocity dispersion for the model, consistent with 
HETG constraints on the gas, and applied the predicted ionic column densities
to fit the absorption-line profiles, using the wavelengths and oscillator strengths from
\citet{bn} and \citet{ven96-lines}. 
The predicted Si{\sc xiv} and Mg{\sc XII} lines
from our ``best-fit'' model (log$U = 1.7$, Log$N_{H} = 22.2$) were 
compared to the {\it Chandra} spectra 
 and showed reasonable agreement between the model and the data.  
Velocity spectra (Fig.~4) indicate a mean bulk velocity 
 -1100 km s$^{-1}$ for the High Ionization gas component.  

While the lower-ionization inner shell lines were expected 
to arise in the UV absorber, 
the signal-to-noise in the data are too poor to allow for 
any detailed profile fitting. 
Therefore, we generated a grid of photoionization models 
around the values suggested by our 
analysis of the LETG spectra \citep{netzer02}. Along with 
the models generated for
the High Ionization component, the tables of ionic column 
densities predicted by Cloudy were used to construct
a grid of high-resolution spectra, including the effects of bound-free 
absorption \citep{vea96}, and resonance and inner shell 
absorption lines \citep{bn}. The spectral grids were used as an 
input to {\tt xspec} in order to fit
the {\it XMM} PN spectra.

\subsection{Spectral fits to {\it XMM} PN data}

Fig.~5 shows the 2001 April and November data relative to 
the powerlaw model (fit across the 3-5 keV band). This illustrates the 
strong curvature in both spectra. 
It is also clear that during November, the source was fainter and more heavily 
absorbed. We investigated the spectral shape for both epochs in light of 
insight from the {\it Chandra} spectral results. 

Prior to fitting we added to the model the set of 
emission features detected in our earlier analysis of the RGS 
data \citep{turner03}, which we
suggested arose in gas outside the absorbers. Further, 
 we discovered two 
 emission features we had not picked up during our RGS analysis, one at 
$\sim$ 0.83 keV, which is likely 
an unresolved blend of Fe L-shell and high order O{\sc viii} resonance 
lines, and another at $\sim$ 1.23 keV, which 
is most likely the Ne~{\sc ix} RRC, for which we measured a width 
corresponding to  
$kT =$ 7.8 x 10$^{-2}$ keV. We used the {\tt xspec} models {\tt zgauss} and 
{\tt redge}  to parameterize these features in the model. 
We  also included  emission 
from the K-shell of Fe, the width  was fixed to be 
$\sigma =13$ eV and energy fixed 
at 6.40 keV, as determined from HEG fits (see later). 

{\it BeppoSAX} data overlapped the {\it XMM}  
observations. The {\it BeppoSAX} data 
showed  the spectrum of NGC~3516 to  
steepen  to 
the canonical $\Gamma \sim 2$ in the 10-200 keV regime covered by the PDS 
spectra. This is 
indicative of the flatness being due to (unmodeled) complex 
absorption or a high-energy cut-off to a flat powerlaw. In light 
of HETG detections of numerous absorption lines, we followed the 
former scenario.  We found that when 
fitting {\it BeppoSAX} data simultaneously with PN data in {\tt xspec}, the 
PN data almost entirely determined the fit parameters by 
statistical domination.  Thus, to best estimate  the strength of any 
reflection component relative to the primary spectrum, we used  
 predominantly PN data from 3-10 keV, tying the line strength to that of 
the reflector. In this way 
we obtained an estimate of $R \sim 1$ for both epochs,  representing 
reflection from 2$\pi$ steradians of material. This value was 
consistent with the spectral curvature observed in the 
{\it BeppoSAX} spectra. Therefore we included the reflected component 
fixed at $R=1$ in all subsequent spectral fits.   
 
We constructed a  model, using 
the parameters derived from the 
absorption lines for the High Ionization component and the \citet{netzer02}
model for the UV absorber, compared to the April EPIC spectrum. 
The model 
drastically underpredicted the absorption at $\gtrsim$ 2 keV and  
 was unable to 
replicate the spectral curvature. Hence, there must be an  
additional large column of
intervening gas (which we will refer to as the ``Heavy'' component). 
However, there did not 
seem to be any excess absorption in the soft band, which 
suggests that the Heavy component is
in a high state of ionization (log$U \gtrsim 1$) and 
that it may not fully cover the continuum source.   
 
Including this third (Heavy) 
component, we refit the April spectrum. In doing so, 
we assumed a priori that the UV 
component fully covers the X-ray continuum source, as suggested 
by \citet{netzer02} 
and \citet{kraemer02}, as does the High Ionization component, as required by 
our absorption line fits.  However, 
we let the covering factor of the Heavy absorber 
component be a 
free parameter. The radial velocity of the High Ionization
component was 
fixed at -1100 km s$^{-1}$, as suggested by the 
observed energy of the H- and He-like absorption lines detected 
in the HETG data. The UV absorber was assumed to have 
an outflow velocity $-$ 200 km s$^{-1}$ (the mean from the strong UV
components). We set the radial velocity of the Heavy component 
to $-$1100 km s$^{-1}$, however
as we will show, the fit was not sensitive to the velocity. 
 We obtained a statistically  
satisfactory fit with the complex model outlined above 
(Fig.~6a).  
The results are summarized in Table~2 (errors are $\chi^2+4.61$). 
 Fig.~6b shows the 
effect of the  Heavy absorber through a model plot  
with that component removed. 

Our spectral fitting returned a solution for Heavy that
 includes a high ionization 
parameter and column density, but a covering factor of only 
50\%, which suggests this exists at 
a small radial distance, since the
X-ray continuum source is $\lesssim$ 1 lt-day in extent. 
The other consequence is that  
at such a low covering fraction, any absorption lines formed 
in Heavy will be undetectable  
since the continuum is dominated by the uncovered fraction. 
Note that the attenuation of the transmitted
continuum by Heavy amplifies this effect. For example, at 
the energies of the high ionization lines
1.5 -- 3 keV, the uncovered continuum contributes 
$\sim$ 75\% of the total flux. Hence, we have
no constraint on the velocity of this component, as no discreet 
absorption lines are detectable with HETG. 

We also tested the effects of screening of the continuum source 
by each of the components, as the photon input to each screen depends 
on any absorption of the nuclear flux by inner layers of gas. 
However, with such a low covering fraction we found little 
effect if the Heavy absorber was positioned closest to the nuclear 
source, with the UV and High Ionization components outside it. 
As this is a plausible radial 
ordering of the absorbers, we proceeded without 
worrying further about the effects of screening on the outer layers. 

The UV component of the model predicts column densities 
for H~{\sc i}, 1.4 $\times 10^{17}$ cm$^{-2}$, C~{\sc iv}, 
4.9$ \times 10^{16}$
cm$^{-2}$, and N~{\sc v}, 1.9 $\times 10^{17}$ cm$^{-2}$, 
which are in agreement 
with the upper limits determined from the STIS spectrum of 2000 October 1, 
 taken when NGC 3516 was in a slightly lower flux state 
\citep{kraemer02}. While Kraemer et al. derived a 
lower limit for S~{\sc iv} of 1.8$ \times 10^{14}$ cm$^{2}$, the UV component 
model predicts an ionic column density only $\sim$ 1/100 as large, which
is consistent with the presence of additional lower 
ionization gas along the line-of-sight to the
nucleus, which does not contribute appreciably to the X-ray opacity. 
The High Ionization component of the model predicts an  H~{\sc i} column
density of 1.4 $\times 10^{14}$ cm$^{-2}$, which may be detectable in Ly$\alpha$ if the absorber
covered the BLR, 
 although the predicted C~{\sc iv} and N~{\sc v} were negligible. 
However, while 
there were components of Ly$\alpha$ at $-$1372 $\pm$9 km s$^{-1}$ and 
$-$994 $\pm$16 km s$^{-1}$ detected in the STIS
spectrum \citep{kraemer02}, the presence of C~{\sc iv} and N~{\sc v} lines at these velocities suggests that the 
High Ionization component is not seen in the UV, hence may not cover the BLR. The fact that the ratios
of the ionic columns to that of H~{\sc i} were anomalously low in these components rules out the possibility
that the Ly$\alpha$ lines arise in the X-ray absorber while the other lines form in lower ionization
gas. 

The model of \citet{netzer02} explains the historic spectra of 
NGC~3516  utilizing intrinsic absorption by the UV absorber, 
whose ionization changes in response to the continuum source. 
We modified the \citet{netzer02} model
to account for the greater complexity in absorption than was evident 
from previous 
datasets and, having achieved a good fit to 2001 April PN data, we 
fit the same model to the 2001 November PN data to determine the cause of the
spectral variations.  
The November data were consistent with no change in column 
density for any absorption component, although there was a 
slight increase in covering factor (Table~2). 
All three absorption zones are consistent with 
 a decrease in
ionization parameter by a factor $1.5$, 
which would be expected as the gas responds to 
changes in illuminating  ionizing radiation. 
Fig.~7 shows 
the best-fitting models to April and November PN data, overlaid, illustrating 
the spectral change.  

Our final model, featuring 
three layers of intrinsic absorption, yielded a 
good fit to the overall shape of the X-ray spectrum. Another satisfactory 
aspect of the fit is that we recovered the canonical underlying 
photon index typical of AGN in the Seyfert class, and 
found a natural explanation for the $\Gamma \sim 2$ slope 
observed above 10 keV in PDS data. This 
gave us an additional level of confidence that we had correctly 
explained the flat observed spectrum of NGC~3516 in this low flux state.

\section{What of the Fe K$\alpha$ Regime?}

\subsection{The Line Core}

Having modeled  the overall spectral shape  we  turned to a detailed 
examination of the residuals around  6.4 keV. 
From HEG data we found that comparison of the core of the Fe K$\alpha$ line 
shows consistent fluxes in  the 2001 April and November epochs, 
expected if the narrow line core is dominated by a contribution from 
the putative distant reflector (Fig.~8). HEG data yielded a core width 
$\sigma=13^{+12}_{-13}$eV, and energy 6.40 keV, as noted previously. 
The PN data, with superior signal-to-noise ratio,  indicated an 
increase in core flux in November versus April, at $>95\%$ confidence. 
If the core were simply velocity-broadened then a 
 width of $\sigma \sim 13$ eV yields a gas velocity  
1400 km/sec. This would occur for gas lying 
$\sim 100$ lt-days from the black hole  
(assumed to have  mass
 $M \sim  2.3 \times 10^7 M_{\odot}$, 
\citealt{wand99}), consistent with the suggestion of variability 
 on timescales of months. However we note a mis-match of 
the {\it XMM}-based model with the HEG data, 
suggestive that the fit to the 
PN spectrum missed some of the core flux in 2001 April spectra. 
Thus the question of core variability remains open.  

\subsection{The Broad Diskline}

A long-standing question has been whether complex absorption 
of the continuum can mimic the broad red wing expected 
from Fe K$\alpha$ contributions from the innermost accretion disk. 
 It is clear  that  with our complex absorption 
model, whatever small excess of flux exists redward of 6.4 keV 
is not obviously  characteristic of a diskline (Fig.~9a,b). 

We attempted to fit the April  ``excess'' using the {\tt xspec} model of 
a diskline following the prescription of \citet{stella} for a 
non-spinning black hole. This model produced an improvement 
$\delta \chi^2 =31$ with inner and outer radii fixed at 6 and 1000 
gravitational radii, respectively,  disk inclination fixed at 
$30^{\rm o}$ and emissivity index at $q=-2.5$. The line energy was 
$6.40 < 6.45$ keV, equivalent width $431^{+193}_{-172}$ eV. 
Thus we conclude the data are consistent with the presence of a substantial 
diskline, although the spectra do not provide compelling evidence for 
such. 

\subsection{Red or blue-shifted narrow components of Fe K$\alpha$ emission}

Turning to the question of narrow line emission, we examined the 
spectra for narrow red or blueshifted Fe k$\alpha$ components, as 
reported previously for this source (\citealt{turner02}, \citealt{bian}). 
 We split the PN data from 2001 April, dividing the spectra  
$\sim 30$ ksec into the 49 ksec exposure, splitting the good interval 
 into two sections, between which the source flux increased by $\sim 12\%$.   

Fig.~9a and b  show the Fe K$\alpha$ profile from 
the two time-selections sampling changes in flux within each epoch of data.  
There is evidence for 
 a very weak feature at rest-energy 6.08 keV, which was 
reported  by \citet{bian}. 
As previously reported for November data \citep{turner02}, 
spectral variability is evident in the Fe K$\alpha$ regime 
down to timescales of a few tens of ksec and these are 
possibly related to changes in continuum flux. 
 While it is tempting to 
attribute the  variable spectral shape in the Fe K$\alpha$ regime 
 (most evident in November) to changes in a broadened line from the 
accretion disk,  our insight 
from HEG spectra leads us to attribute this variability to 
 rapid (tens of ksec or less) changes in the  energy-shifted 
narrow components of Fe K$\alpha$. 
We also note the presence of additional varying 
features not previously reported,
e.g. close to 3 keV (Fig.~9b). While it is tempting to speculate 
on their origin, much longer 
{\it XMM} observations are needed to progress on reliably 
characterizing the line identifications and behaviors.

\section{Discussion and Conclusions}

\subsection{Nature of the Absorbers}

 We find a model for NGC~3516 which has three intrinsic layers of 
absorption with significantly different columns and ionization-states, and, 
when determined, outflow velocities. 
The UV absorber has previously been detected in both UV and soft X-ray 
regimes. A High Ionization absorber is required to produce the strong H- and He-like
Mg, Si, and S lines detected in the 2001 April HETG spectrum. Finally, the Heavy absorber has 
been detected previously but we 
find the first constraint on its covering fraction.  

The first evidence for the Heavy component came from 
{\it Ginga} data taken when NGC~3516 was observed 
several times in flux-states close to that found for 2001 April. 
In their modeling of the complex X-ray absorption detected in these data, one of the solutions suggested
by \citet{kol93} included a warm component which introduced an additional iron edge at
$\gtrsim$ 7.5 keV, with a column of a few $\times 10^{23}\ {\rm cm}^{-2}$.
Using the same dataset, \citet{np94} determined the iron edge energy to be $\sim 7.8$ ke V and   
derived a warm absorber column density similar to that suggested by Kolman et al. The ionization-state
suggested by the {\it Ginga} results is similar to the Heavy component in our models.
Costantini et al (2000) also found the presence of a large 
column of highly ionized gas in {\it BeppoSAX} spectra, although at 
$N_H \sim 2 \times 10^{22} {\rm cm}^{-2}$ this is 
different to the {\it Ginga}
solution and resembles our High Ionization rather than our Heavy component.
Since the High Ionization component would be undetectable at low resolution if
a significantly larger column of gas were present, the discrepancy may indicate that the
Heavy column was absent during the earlier 
{\it BeppoSax} observations. 

Based on our models, we can determine the densities, $n_{H}$, of the High Ionization and
Heavy components as a function of their radial distance, $r$. 
For the High Ionization absorber,
log$U =$ 1.7, hence for 
a luminosity in ionizing photons of 
$\sim 2 \times 10^{53}$ s$^{-1}$ \citep{kraemer02},
$rn_{H}^{1/2} \sim 1.0 \times 10^{20} {\rm cm^{-1/2}}$, 
while for the Heavy absorber, log$U =$ 1.2 and
$rn_{H}^{1/2} \sim 1.8 \times 10^{20} {\rm cm^{-1/2}}$.  
Although we cannot deconvolve the densities and distances of the absorbers,
it is implausible that the physical depth of 
the absorbers ($\Delta r$ $=$ n$_{H}$/N$_{H}$) could be more than
a small fraction of their radial distance. 
For example, for the High Ionization component, if $\Delta r/r$ $=$ 0.1 
then  n$_{H}$ $=$ 2.5 x 10$^{6}$ cm$^{-3}$
and $r = 6.3 \times 10^{16}$ cm, which is 
approximately equal to the lower limit derived for the
UV absorber \citep{kraemer02}. The same line of 
reasoning suggests that the Heavy component lies somewhat
closer to the central black hole, which is consistent with partial covering. 
In fact, since the size of the X-ray continuum source is $<$ light-day, one might expect  the radial distance
of the Heavy absorber to be of the same order of magnitude. 
Alternatively, partial covering may result if the 
absorber is patchy or filamentary.  

Although we cannot constrain the densities of the Heavy and UV components, the ratios of their mean electron temperatures 
to their ionization parameters
can be used to estimate 
their relative thermal pressure. The model predicts T $=$
10$^{5.63}$ K, 10$^{5.99}$ K and 10$^{4.44}$ K for the Heavy, High Ionization, and UV absorbers, respectively.
The thermal pressure of the UV absorber is  $\sim$ 6  times that of the Heavy absorber and 9 times that of
the High Ionization absorber. Hence, these three components cannot be in 
thermal equilibrium with respect to each other. To examine the thermal stability of the
absorbers, in Fig.~10   
we show the relationship of T and the so-called ``pressure'' 
ionization parameter U/T (generally
known as an S-Curve). As discussed most recently in 
\citet{kk}, there are two regions 
where a photoionized gas is stable to thermal perturbations: at low ionization/temperature, when
line cooling is efficient, and high ionization/temperature, when thermal balance is achieved via
Compton processes. In the intermediate region, the gas can be unstable if $dT/d(U/T) < 0$. There is no such
region for our models, in part due to the additional cooling provided by the over-abundance of nitrogen (in contrast,
see the S-curve in \citealt{kraemer02}). Nevertheless, as a result of the large changes in temperature which result from
small changes in $U$, the intermediate region is only marginally thermally stable (for a more detailed
discussion, see \citealt{kk}). 
While the UV absorber is within the line-cooled region, the two highly
ionized components lie in the intermediate region, and should undergo large changes in temperature in response to variations in the 
incident continuum.

Given the high ionization state of the High Ionization and Heavy absorbers, they are quite 
transparent to the incident continuum, with the main source of opacity
being bound-free transitions from highly ionized O, Ne, Mg, Si, S, and Fe.
Our models predict an average force multiplier, the ratio 
of the radiation pressure force from all absorption and scattering process to
that caused by Compton scattering, $\sim$ 3. As a consequence, the central source
would have to be radiating at roughly 30\% of its Eddington Luminosity in order to
generate a radiatively driven outflow. Also, our models predict mean temperatures
which are too low for a thermal wind to achieve such 
a high velocity \citep{bals93}. This suggests the possibility that these 
absorbers are associated 
with an MHD wind (e.g., \citealt{emm92}). 
As noted above, their high ionization state strongly suggests an
origin close to the continuum source and \citet{emm92} predict that the outflow velocity in an MHD flow
should increase as the radial distance of the launch pad decreases. Furthermore, based on their
models of MHD flows in the BLR gas in NGC 5548, \citet{bott97a} predict line-of-sight
velocities of the same order as that observed for the High Ionization absorber. 
Finally, as discussed above, these absorbers are only marginally thermally stable, although they appear to have been
present in earlier observations. If the absorbers were permeated by magnetic fields of sufficient strength
(i.e., on the same order as their thermal pressure), the magnetic pressure would act to 
stabilize them \citep{bott00}. The presence of internal magnetic fields would also suggest an MHD origin.

In principal, given the line-of-sight distance of the absorber 
from the black hole, the line-of-sight velocity of the material, the disk 
orientation with respect to the observer, and the black hole mass,  
 it may be  possible to constrain parameters of a 
\citet{BP82} type MHD disk wind model applied 
to NGC 3516. This is the subject of future research efforts. We note that 
an early attempt to constrain \citet{BP82} 
model parameters in NGC 5548, using its 
BLR C{\sc iv} emitting region, is described in \citet{bott97b} and was 
relatively successful in describing emission line shape and variability. 
This model was extended to the UV/X-Ray warm absorber of NGC 5548 but 
analysis was hampered by the then relatively low resolution  
X-Ray data of the pre-{\it Chandra}, pre-{\it XMM} era \citep{bott00}. We also 
note that the masses of the black holes in NGC 3516 and NGC 5548 are 
similar. It would therefore be tempting to directly apply the results of 
the NGC 5548 model to NGC 3516. We caution however that slight differences 
in \citet{BP82} MHD flow parameters and/or orientation can lead to large 
differences in the radial velocity and line-of-sight position of absorbing 
material.

\subsection{The Fe Emission}

We note with interest the 
remarkable similarity between the Fe K$\alpha$ profile of NGC~3516 and 
that of Mrk~766 (\citealt{p03},\citealt{turner03}). Mrk~766 and NGC~3516 
are two of very 
few  Seyfert galaxies known to 
exhibit narrow and highly-redshifted 
Fe emission lines 
(\citealt{turner02}, \citealt{turner03}).   
Another similarity 
is that Mrk~776 also shows evidence for a large column of ionized gas 
\citep{p03} with $N_H \sim 2 \times 10^{23} {\rm atom\ cm^{-2}}$. 
If the emission is from ionized Fe then it would be natural to expect 
the heavy absorption and emission from ionized Fe
 to be observationally linked. 
It may be that we can only see the weak narrow Fe lines in sources 
with the largest columns of circumnuclear gas. 
If the narrow redshifted  Fe emission lines \citep{turner02} do 
originate in the 
highly-ionized gas components then previous speculations will have 
underestimated their velocity. The  
emission from the K-shell of Fe in NGC~3516 is expected to peak at 
Fe {\sc xx} ($\sim 6.51$ keV), perhaps  raising estimates of the 
 outflow  velocity for the emitting gas 
responsible for the shifted features reported by 
\citet{turner02} up to $\sim 0.16c$, 
much higher than outflow velocities estimated for  the absorption lines 
reported here.  
 
Other AGN have shown evidence for zones of high-ionization gas, 
e.g. NGC ~3783 \citep{reeves04} which shows an Fe resonant absorption 
line, where again, the inclusion of the absorber significantly reduces 
the inferred equivalent width of any broad component to the diskline. 
Even higher column ($\sim 10^{24}\ {\rm atom\ cm^{-2}}$)  
outflows of gas have also been suggested with outflow 
velocities up to 0.1$c$ in QSOs such as 
PG~1211$+143$ \citep{pounds03}, 
 PDS 456 \citep{reeves03} and the NLSy1  
NGC 4051 \citep{pounds04}. In those cases the column 
density is high enough that absorption edges + Fe K-shell resonant 
absorption lines can be detected.  

While current data leave us tantalizingly close to answering the question 
as to whether high-column, high-ionization absorbers are 
masquerading as broad disklines, a conclusive answer will have to wait until 
long exposures are taken on AGN using the microcalorimeter on 
{\it ASTROE-2}.

\section{Acknowledgments}

We are grateful to Luigi Piro and the {\it BeppoSAX} satellite 
operation team for 
arranging simultaneous {\it BeppoSAX} coverage of NGC~3516 at short 
notice, to cover the  {\it XMM} observations.    We also thank the 
{\it XMM} and {\it Chandra} satellite operation teams. 
Thanks to Michael Crenshaw for useful discussions.   
T.J.\ Turner acknowledges support from NASA 
grant GO1-2099B.  

\clearpage

\begin{deluxetable}{lccc}
\tablewidth{0pc}
 \tablecaption{
  X-ray Flux}
 \tablehead{
 \colhead{Epoch} &
 \colhead{2-10 keV\tablenotemark{a}} &
\colhead{0.5-10 keV\tablenotemark{a}} &
\colhead{7 keV\tablenotemark{b} } }
 \startdata
\multicolumn{4}{c}{Observed fluxes} \\
\hline
2000 Oct & 1.04 & 1.23 & 0.78 \\
2001 April & 2.27 & 2.70 & 1.72 \\
2001 Nov & 1.63 & 1.92 & 1.35 \\
\hline
\multicolumn{4}{c}{Absorption-Corrected Fluxes} \\
\hline
2000 Oct & 1.42 & 2.17 & 0.82 \\
2001 April & 2.95 & 4.75 & 1.82 \\
2001 Nov & 2.22 & 3.39 & 1.45 \\
\hline
\multicolumn{4}{c}{Absorption-Corrected PowerLaw Flux} \\
\hline
2000 Oct & 1.21 & 1.95 & 0.61 \\
2001 April & 2.72 & 4.50 & 1.62 \\
2001 Nov & 1.89 & 3.05 & 1.15 \\
\hline

		\enddata

\tablenotetext{a}{Fluxes as ergs cm$^{-2}$ s$^{-1}$ in units of 10$^{-11}$.}
\tablenotetext{b}{Fluxes as keV cm$^{-2}$ s$^{-1} {\rm keV}^{-1}$ in units of 
10$^{-3}$.}

\end{deluxetable}

\clearpage

\begin{deluxetable}{lcc}
\tablewidth{0pc}
 \tablecaption{PN fit parameters} 
 \tablehead{
\colhead{} &
 \colhead{2001 April} &  \colhead{2001 Nov}  }
 \startdata

$\Gamma$ & 1.82$\pm0.01$ & 1.77$\pm0.02$ \\

Cov \tablenotemark{a} & 44$\pm6$\% & 58$\pm5$\% \\

Log N$_{H,UV}$ & 21.85$\pm0.15$  & 21.60$\pm0.15$  \\
Log U$_{UV}$  & $-$0.80$\pm0.02$  & $-$0.99$\pm0.04$ \\

Log N$_{H,hi}$ &22.2$f$\tablenotemark{b} & 22.2$f$ \\ 
Log U$_{hi}$ & 1.7$f$ & 1.5$f$ \\

Log N$_{H,heavy}$ &23.40$^{+0.03}_{-0.05}$  & 23.40$\pm0.04$ \\ 
Log U$_{heavy}$ & 1.21$^{+0.02}_{-0.04}$ & 1.09$^{+0.01}_{-0.04}$ \\ 

$E_{FeK}$ (keV) &  6.40$f$ & 6.40$f$ \\
$n_{FeK}$ \tablenotemark{c} & 3.64$^{+0.38}_{-0.38}$ 
	& 4.91$^{+0.22}_{-0.36}$ \\

$\chi^2\ /d.o.f.$ & 1801/1521 & 3248/2770 \\
		\enddata
\tablenotetext{a}{Fraction of nuclear source covered by the heavy absorber}
\tablenotetext{b}{$f$ denotes that a parameter was fixed in the fit} 
\tablenotetext{c}{Photons in the line in units 10$^{-5}$ }

\end{deluxetable}

\clearpage

\typeout{FIGS}

\begin{figure}
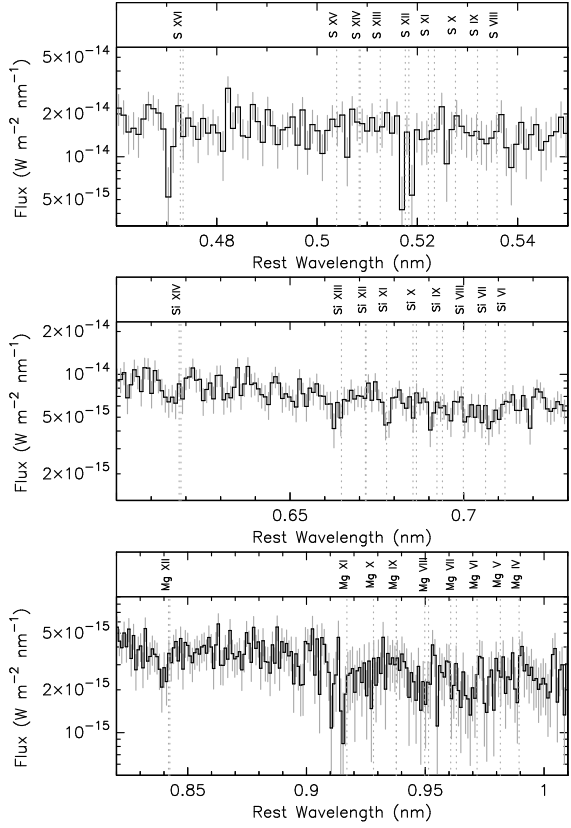
   
         \epsscale{0.7} 
	 \includegraphics[scale=0.3,angle=270]{f1a.eps}\\
	 \includegraphics[scale=0.3,angle=270]{f1b.eps}\\
	 \includegraphics[scale=0.3,angle=270]{f1c.eps}         
	        \caption[Absn lines]{
The panels show key sections of HETG data, where we have 
combined HEG and MEG data 
from the $\pm1^{\rm st}$-order spectra. 
The most prominent absorption lines are evident in the different panels, with  
a) S{\sc xvi}; b) Si{\sc xiv} and c) Mg{\sc xii}. The dotted lines show where 
those and other absorption lines would be expected.
We have ignored some artifacts of the instrument which 
mimic absorption features;  for example,  
the Mg {\sc ix} region falls on the S3/S4 chip gap in the positive 
first order spectrum, producing a depression in the observed spectrum. }
\end{figure}

\begin{figure}
\epsscale{0.75}
\includegraphics[scale=0.60,angle=270]{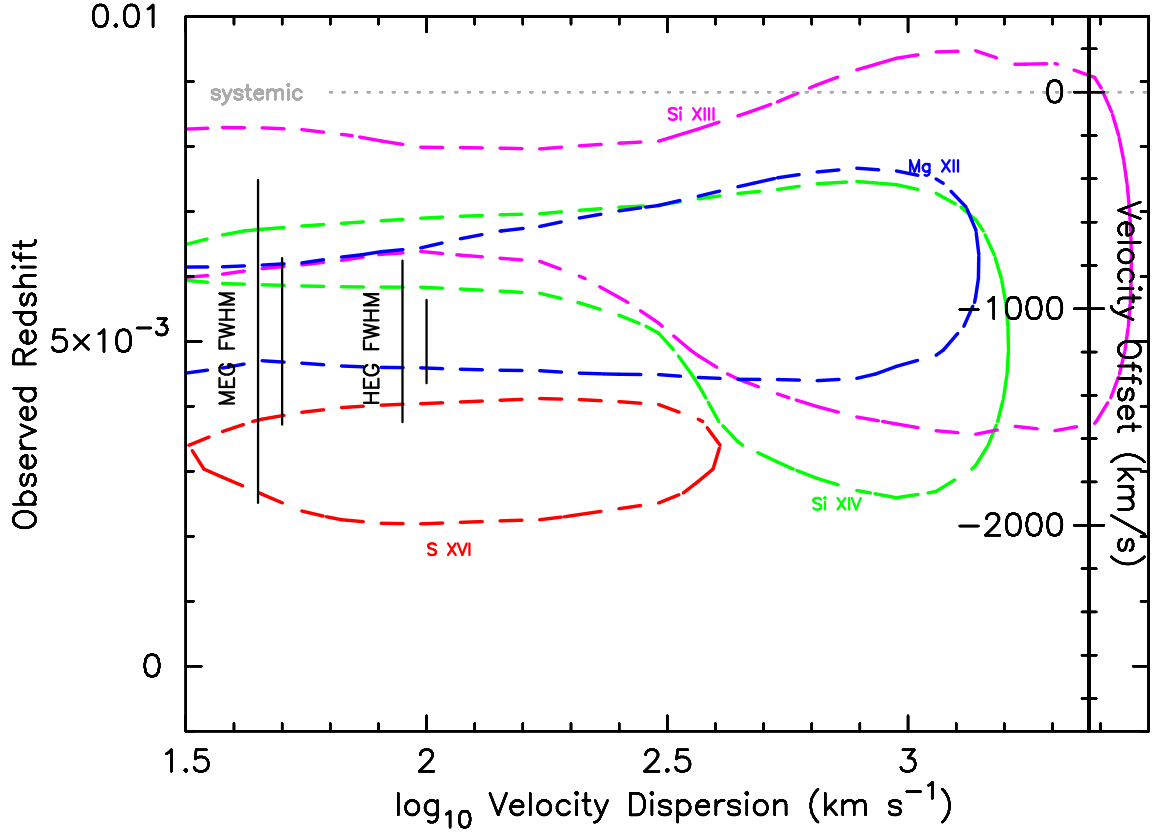}
\caption{ 
Velocity dispersion of the absorber versus 
outflow velocity. The 
90\% confidence contours are shown 
for the kinematic parameters of the lines from the HETG data. 
The absorption lines are 
blueshifted  relative to the systemic redshift, with S{\sc xvi} 
showing a higher blueshift than the other lines, as discussed in 
the text. It is also evident that the lines yield no significant 
constraint on the velocity dispersion of the absorbing gas, and so 
a value of 300 km s$^{-1}$ has been adopted. 
The FWHM for the MEG and HEG are shown for reference for the lowest and 
highest wavelengths (these are for S{\sc xvi} and Mg{\sc xii} lines).  
  }  
\end{figure}

\begin{figure}
\epsscale{0.75}
\includegraphics[scale=0.7,angle=270]{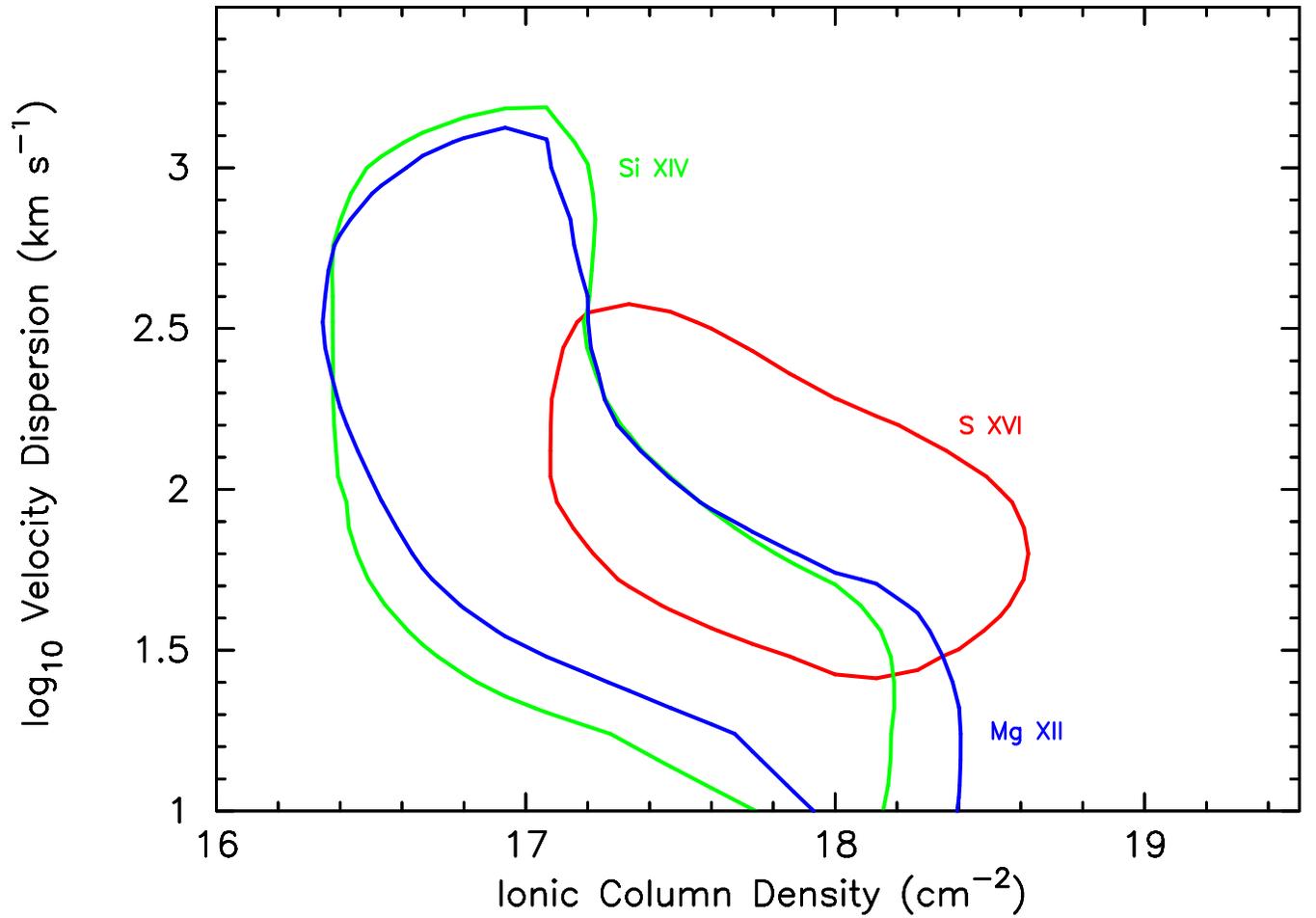}
\caption{The 90\% confidence contours for the H-like absorption lines from 
  S, Si and Mg. Contours were calculated assuming 
two interesting parameters (those plotted). 
The gas was assumed to cover the nuclear source 
completely and the outflow velocity of each line was left to float in the fit.  } 
\end{figure}

\begin{figure}
\epsscale{0.75}
\includegraphics[scale=0.7,angle=270]{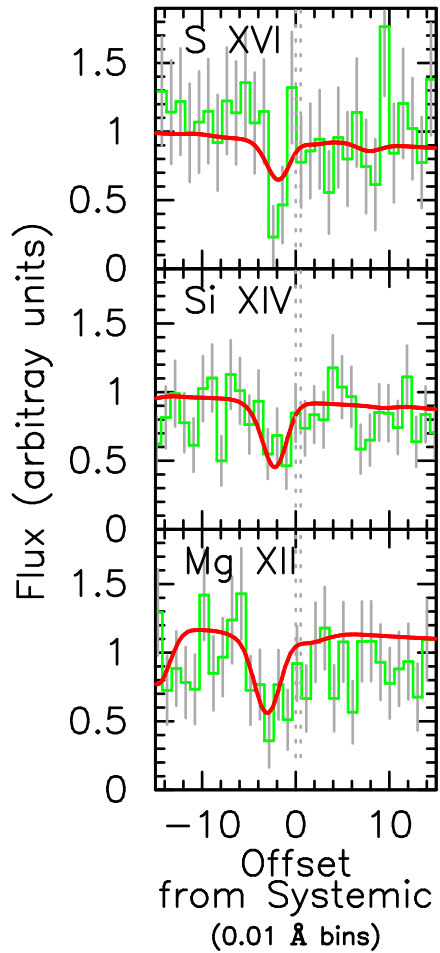}
\caption{Velocity spectra, binned at 0.01\AA\  for the three strongest 
lines detected in the 2001 April MEG (only) spectra from obsid 2080. 
The red line shows the 
model fit to the absorption feature convolved with the MEG 
spectral resolution  and  assuming a velocity dispersion 
of 300 km s$^{-1}$.  } 
\end{figure}

\begin{figure}
\epsscale{0.75}
\includegraphics[scale=0.5,angle=270]{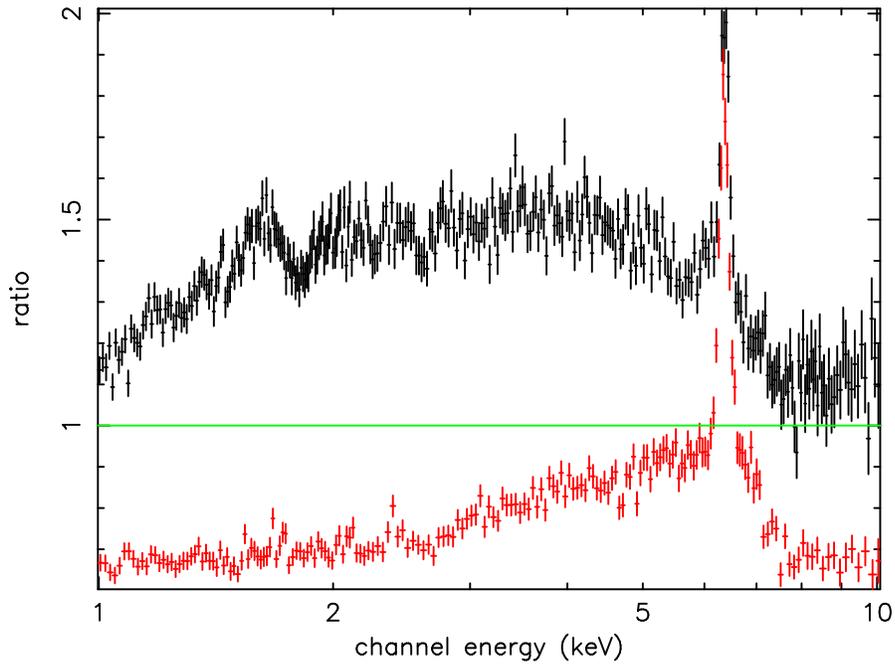}
\caption{Ratio of PN data from 2001 April (top line, black) 
and November (bottom line, red) to a model which represents the 
mean power-law continuum of the two epochs fit over in 
the 3-10 keV regime. This plot illustrates the flattening of the spectrum 
in November compared to April, due to the drop in ionization-state of 
the absorber as the source flux decreases. }  
\end{figure}

\begin{figure}
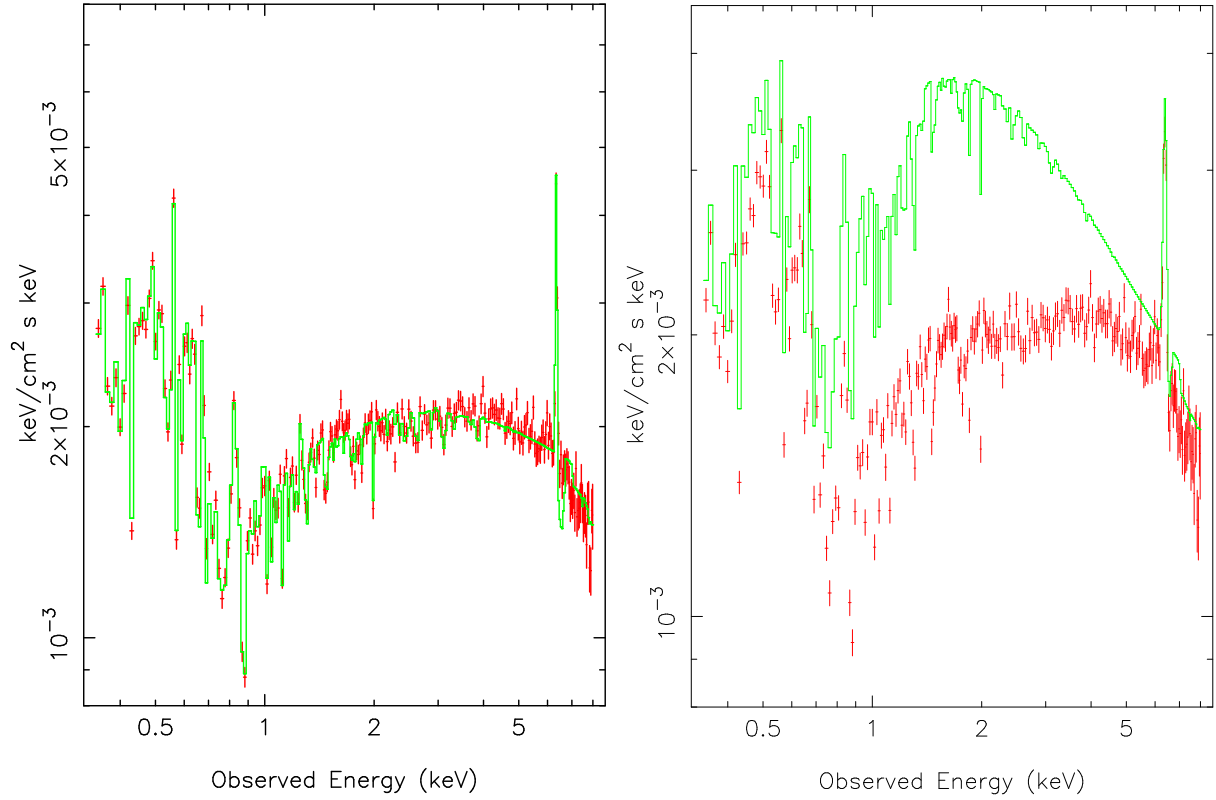

\epsscale{0.75}
\includegraphics[scale=0.5,angle=0]{f6a.eps}
\includegraphics[scale=0.5,angle=0]{f6b.eps}
\caption{The 2001 April PN data versus  
a) the best-fitting model 
including three intrinsic absorbers and 
b) the fit with the Heavy Component
 removed from the model to illustrate the degree of curvature 
attributed to that zone. }  
\end{figure}

\begin{figure}
\epsscale{0.75}
\includegraphics[scale=0.5,angle=0]{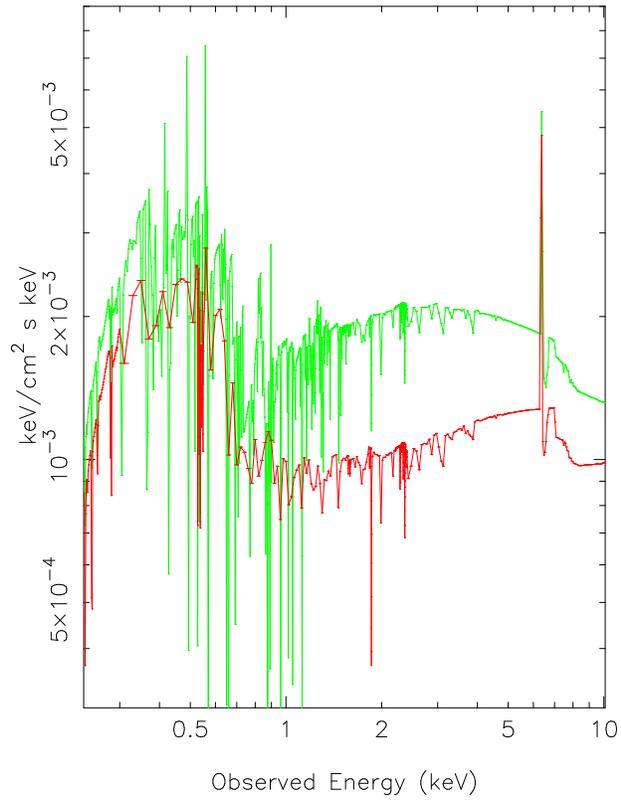}
\caption{The unfolded spectrum (from fits to the EPIC PN data) compared to the 
model discussed in sections 4 and 5. The green line (top) represents 
the 2001 April data while the red line (bottom) is based upon 2001 
November data.}
\end{figure}

\begin{figure}
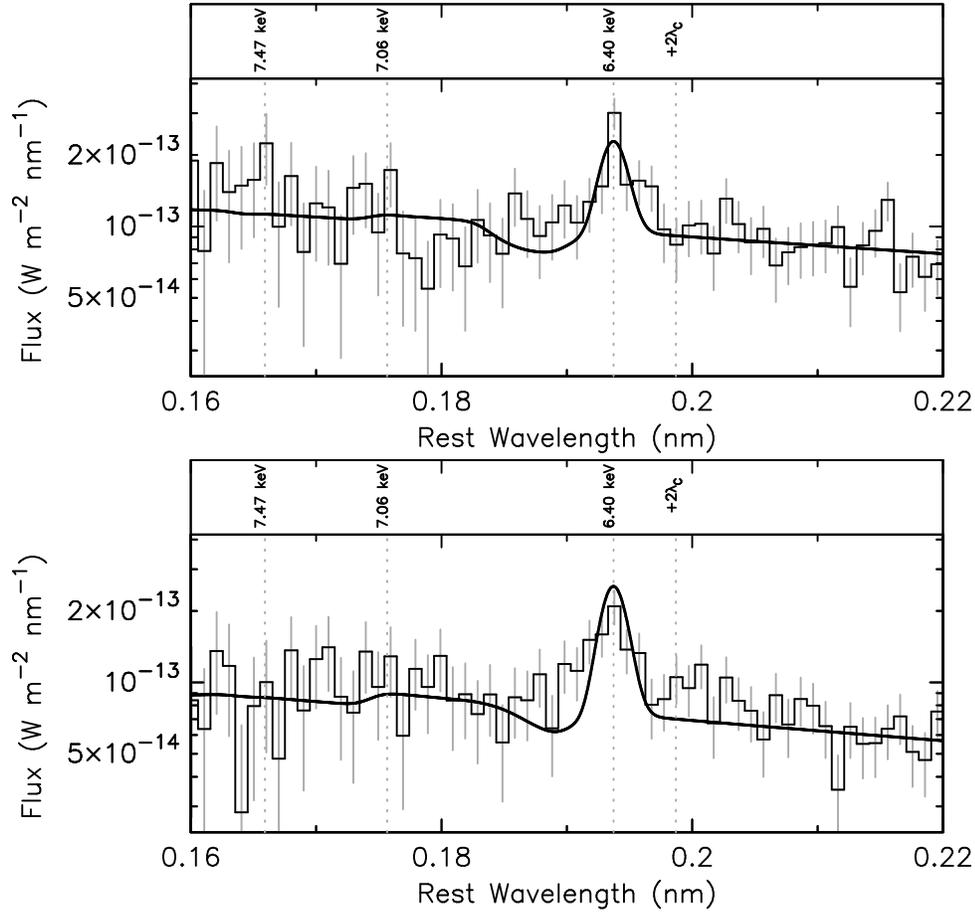

\epsscale{0.75}
\includegraphics[scale=0.5,angle=270]{f8a.eps}
\includegraphics[scale=0.5,angle=270]{f8b.eps}
\caption{
Comparison of the core of the Fe K$\alpha$ line 
detected by the HEG for 
a) the 2001 April and b) November epochs. Positive and 
negative parts of the first order have been combined in each case. }
 \end{figure} 

\begin{figure}
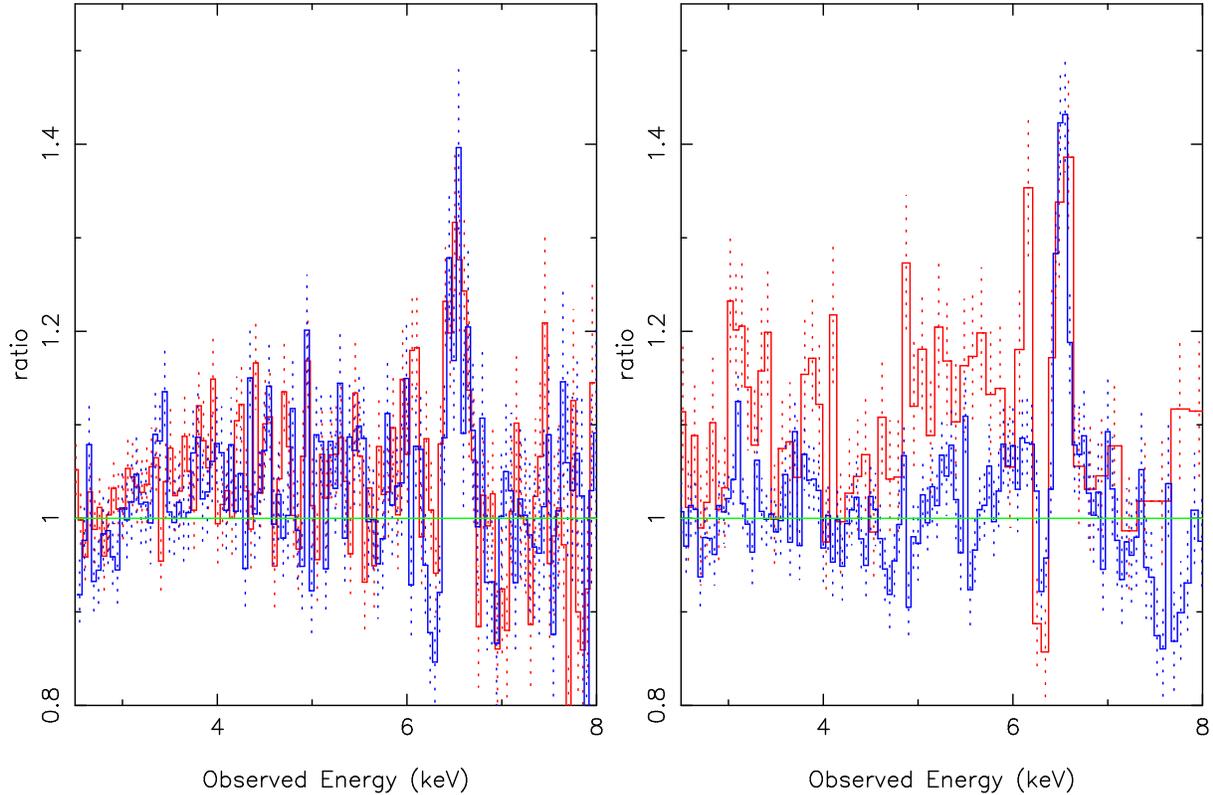

\epsscale{0.75}
\includegraphics[scale=0.5,angle=0]{f9a.eps}
\includegraphics[scale=0.5,angle=0]{f9b.eps}
\caption{Ratio of data/model for our best-fitting model from 
Table~2. Each of the PN datasets have been 
split in time to sample different flux states, as discussed 
in the text. The red line represents the higher flux state in 
each case. Rapid changes evident in the profile during November 
are due to rapid variations in the flux of narrow and redshifted Fe lines 
\citep{turner02}. April data do not allow us to sample such a wide range of 
flux and while we see evidence for a previously-reported line 
at $\sim 6.1$ keV \citep{bian} no significant variability is evident 
within the good section of PN data. 
} 
\end{figure}

\begin{figure}
\epsscale{0.75}
\includegraphics[scale=0.5,angle=90]{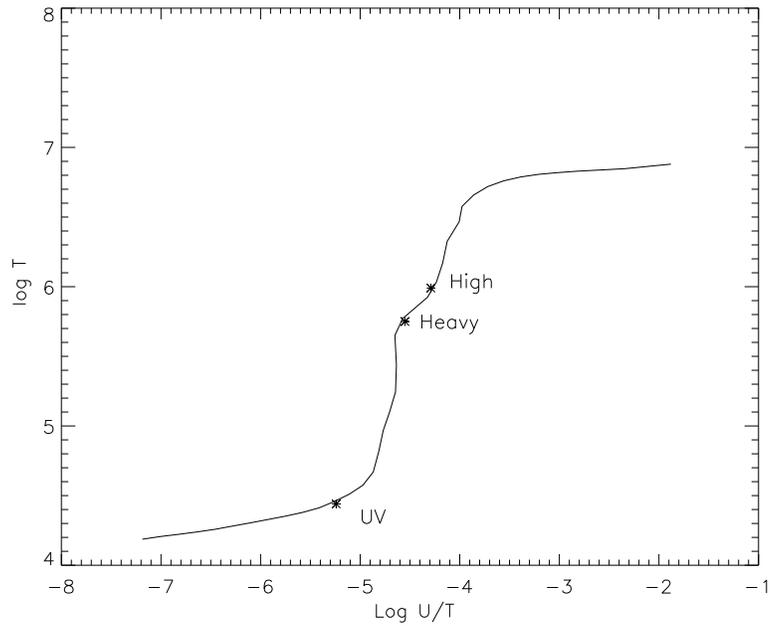}
\caption{
Thermal stability curve for models generated with our assumed SED. The flat 
sections, at low and high temperature, are the stable line-cooled and 
Compton-cooled regions, respectively. Note that, while there are no thermally 
unstable ($dT/d(U/T) < 0$) regions, in the region where the curve is nearly 
vertical small changes in ionization 
can cause large changes in temperature. The 
locations of our three models for the 2001 April spectra are shown.}
\end{figure}

\end{document}